\documentclass[sigconf,natbib=false,nonacm]{acmart}

\RequirePackage[
  datamodel=acmdatamodel,
  style=acmnumeric,
  sorting=none
  ]{biblatex}
  
\usepackage{algorithm}
\usepackage{algpseudocode}

\usepackage{tabularx}
\usepackage{booktabs}
\usepackage{multicol}
\usepackage{multirow}
\usepackage{soul} 
\usepackage{xcolor, colortbl} 
\usepackage{changepage,threeparttable} 
\usepackage[rgb,dvipsnames]{xcolor}

\definecolor{rowgray}{gray}{0.9}

\newcommand{\hlc}[2][Cyan]{{%
    \colorlet{foo}{#1}%
    \sethlcolor{foo}\hl{#2}}%
}

\newcommand{\highlighted}[1]{
\hlc[rowgray]{#1}
}

\usepackage{stfloats}

\hypersetup{
    colorlinks=true,
    linkcolor=Blue,
    filecolor=magenta,      
    urlcolor=Blue,
    citecolor=Blue,
}
\usepackage{wrapfig}

\graphicspath{{Figures/}}

\usepackage[many]{tcolorbox}    	
\definecolor{main}{HTML}{31363F}    
\definecolor{sub}{HTML}{EEEEEE}     
\tcbset{
    sharp corners,
    colback = white,
}  
\newtcolorbox{boxH}{
    colback = sub, 
    colframe = main, 
    boxrule = 0pt, 
    leftrule = 5pt 
}
\newtcolorbox{boxA}{
    fontupper = \bf,
    boxrule = 1pt,
    colframe = main 
}

\newtcolorbox{boxK}{
    sharpish corners, 
    boxrule = 0pt,
    toprule = 4.5pt, 
    enhanced,
    fuzzy shadow = {0pt}{-2pt}{-0.5pt}{0.5pt}{black!35} 
}
\usepackage{subfiles} 
\begin{document}


\title{Inside Out: Uncovering How Comment Internalization Steers LLMs for Better or Worse} 

\author{Aaron Imani}
\email{aaron.imani@uci.edu}
\orcid{0000-0001-7183-5468}
\affiliation{%
  \institution{University of California, Irvine}
  \city{Irvine}
  \state{CA}
  \country{USA}
}

\author{Mohammad Moshirpour}
\email{mmoshirp@uci.edu}
\orcid{0009-0009-9763-0124}
\affiliation{%
  \institution{University of California, Irvine}
  \city{Irvine}
  \state{CA}
  \country{USA}
}

\author{Iftekhar Ahmed}
\email{iftekha@uci.edu}
\orcid{0000-0001-8221-5352}
\affiliation{%
  \institution{University of California, Irvine}
  \city{Irvine}
  \state{CA}
  \country{USA}
}

\thanks{Accepted in the 48th IEEE/ACM International Conference on Software Engineering (ICSE)}

\renewcommand{\shortauthors}{Imani et al.}

\begin{abstract}
While comments are non-functional elements of source code, Large Language Models (LLM) frequently rely on them to perform Software Engineering (SE) tasks. Yet, where in the model this reliance resides, and how it affects performance, remains poorly understood. We present the first concept-level interpretability study of LLMs in SE, analyzing three tasks - code completion, translation, and refinement - through the lens of internal comment representation. Using Concept Activation Vectors (CAV), we show that LLMs not only internalize comments as distinct latent concepts but also differentiate between subtypes such as Javadocs, inline, and multiline comments. By systematically activating and deactivating these concepts in the LLMs' embedding space, we observed significant, model-specific, and task-dependent shifts in performance ranging from -90\% to +67\%. 
Finally, we conducted a controlled experiment using the same set of code inputs, prompting LLMs to perform 10 distinct SE tasks while measuring the activation of the comment concept within their latent representations.
We found that code summarization consistently triggered the strongest activation of comment concepts, whereas code completion elicited the weakest sensitivity. These results open a new direction for building SE tools and models that reason about and manipulate internal concept representations rather than relying solely on surface-level input. 

\end{abstract}




\maketitle

\section{Introduction}

Large Language Models (LLM) are increasingly used to automate diverse Software Engineering (SE) tasks \cite{fan2023, liu2024largelanguagemodelbasedagents}, such as comment generation \cite{geng2024multi-intent}, code translation \cite{pan2024lost}, and code refinement \cite{guo2024code-refinement}. In these tasks, LLMs typically take source code as input, which broadly consists of two components: code and code comments (hereafter, comments). While code defines program functionality, comments are natural language annotations that convey intent \cite{chen2021codecomment}, explain logic, or provide contextual information. Comments play a vital role for humans when understanding and maintaining software \cite{xia2018pc, jabrayilzade2024taxonomy, gene2022forester}, supporting tasks such as identifying self-admitted technical debt \cite{FARIAS2020} and predicting bugs \cite{radmanesh2024investigatingimpactcodecomment}. 

Due to comments' distinct nature, recent studies have examined the effect of including comments in the input on LLM performance \cite{imani2025omega, macke2024testing, nikiema2025codebarrierllmsactually, hossain2025doc2oracllinvestigatingimpactdocumentation, liu2024repoqaevaluatinglongcontext}. 
However, prior research in Natural Language Processing (NLP) has shown that LLM behavior depends not only on the input prompt but also on the model's internal representation space \cite{turner2024steeringlanguagemodelsactivation, gao2025evaluatebiasmanualtest}, where its learned semantics are encoded. For instance, Xu et al. \cite{xu2024safetyrisks} demonstrated that the concept of ``prompt safety’’ is \textit{internalized} within LLMs; prompts containing this concept can be linearly separated from those without it in the models’ internal representations. By suppressing this concept in the latent space, they induced models to respond to malicious prompts with a higher success rate than previous approaches that relied solely on prompt-level manipulations.
These findings suggest that while changes to input, such as the inclusion or omission of comments, can influence LLM behavior, the learned concepts embedded in the model's internal representations are equally, if not more, important. This motivates the need to investigate whether comments are internalized as a concept in LLMs and how this internalization affects model behavior when processing source code. Such understanding is essential not only for interpreting LLM behavior in SE tasks but also for guiding the design of prompts, training data, or interventions that aim to leverage comments more effectively.
To the best of our knowledge, no prior work has examined whether LLMs internally represent or rely on comments as part of their learned representations, nor how such internalization affects their performance in downstream tasks. Moreover, addressing this gap is crucial since existing studies confirm that comments influence LLM behavior, yet it remains unclear whether this influence arises from prompt-level cues or from a deeper internalization of comments as conceptual entities within the model.


To investigate this internalization, we require an interpretability method that meets the following criteria: \textcircled{1} It should provide global explanations \cite{zhao2024explainability}, allowing us to understand what LLMs have learned internally, rather than offering case-specific justifications for individual outputs. \textcircled{2} Since comments are high-level, human-interpretable abstractions, the approach must operate at the concept level rather than relying on uninterpretable low-level features, for example, instead of isolating the effect of a single token or a particular neuron activation, it should assess whether the model has learned a coherent, abstract notion of  ``code comment''. \textcircled{3} The approach must be scalable and computationally efficient to enable analysis across diverse prompts, tasks, and LLMs.

Among concept-based interpretability methods that offer global explanations, recent techniques such as circuit tracing \cite{ameisen2025circuit} have gained traction. However, these approaches are computationally intensive-requiring hundreds of GPU hours even for models with fewer than 10 billion parameters\cite{ameisen2025circuit}, which violates our scalability criterion \textcircled{3}. To address this limitation, we adopt a more lightweight yet effective interpretability technique called Concept Activation Vectors (CAV) \cite{kim2018cav}. 

CAV offers a framework for interpreting the internal representations of neural networks with respect to human-defined concepts. The method is input-agnostic, provides global insights into the presence of concepts across model layers \cite{gao2025evaluatebiasmanualtest, xu2024safetyrisks}, and is highly scalable, requiring only the training of a linear classifier on a small dataset \cite{kim2018cav, xu2024safetyrisks}. 
Given a concept and two sets of examples, one containing the concept and one without it, CAV propose that if a linear classifier can accurately distinguish between these two sets, the classifier's output can be used to measure the degree to which that concept is activated in unseen data \cite{xu2024safetyrisks}. Specifically, for a new input, a prediction probability close to 1 indicates strong \textit{activation} of the concept in the model’s internal representation, whereas a probability close to 0 indicates that the concept is \textit{deactivated}.
Therefore, given the properties of CAV (its global scope, concept-level reasoning, and both data and computational efficiency), it fulfills all three of our criteria for a suitable interpretability approach.



Since CAV was originally introduced for deep neural networks prior to the rise of transformer architectures \cite{vaswani2017attention} and LLMs, it is important to validate their applicability in this newer context. The assumptions and mechanisms underlying CAV-such as linear separability of concept representations in the activation space-must remain meaningful when applied to the high-dimensional, multi-layered representations learned by LLMs. Encouragingly, recent research in NLP has successfully applied CAV to LLMs for various purposes which includes evaluating social and demographic biases \cite{gao2025evaluatebiasmanualtest}, steering responses through targeted activation manipulation \cite{turner2024steeringlanguagemodelsactivation}, and exposing safety vulnerabilities by analyzing concept deactivation effects \cite{xu2024safetyrisks}. These examples provide empirical support for using CAV in the LLM setting and motivate their use in our investigation on internalization of comment as a concept.


To analyze the potential internal reliance on comments as a distinct concept, the first step is to determine whether they are internalized in LLMs' latent space. Therefore, we formulate our first research question as below:

\noindent\textbf{RQ1: Do LLMs internally recognize comments as a concept?}

Following the common practice in SE, we focus on the Java programming language due to its widespread industrial use \cite{TIOBETIOBE} and its extensive adoption in SE research \cite{li2024omg, wang2023commit-issue}. In Java, as in many other languages, comments appear in various forms depending on their scope and level of detail. 
Javadocs are used to document classes and methods at a high level, while inline comments are brief, single-line annotations that clarify one or a few lines of code within method bodies. Additionally, developers often write sequences of consecutive inline comments-referred to as multiline comments in this work-to provide more detailed, localized explanations of complex logic \cite{freitas2012comment-analysis}. Due to these distinctions, prior work on comment analysis has often examined comment types separately to understand their unique roles \cite{freitas2012comment-analysis, huang2023methodcomment}. Building on our first research question, we investigate whether LLMs internalize each of these comment types as distinct concepts leading to our second research question:

\textbf{RQ2: Do LLMs internalize different types of comments as distinct concepts?}


This analysis helps us understand whether LLMs treat different comment types uniformly or assign varying functional importance to each. For example, if an LLM relies more heavily on multiline comments than Javadocs for specific tasks, this could inform how to write or prioritize comments for LLM-centric workflows. It also enables us to identify which comment types are most influential in shaping the model’s internal representations and behavior-insights that can guide training objectives and prompt engineering practices.

To maximize the generalizability of our results, we conducted experiments on three distinct types of state-of-the-art LLMs: a code LLM, a general-purpose LLM, and a reasoning-oriented LLM. Across all three models, we found strong evidence that the concept of comments has been internalized with high accuracy. However, the extent of internalization varied across comment types, with Javadocs emerging as the most robustly learned. Given the recognition of the comment as a concept in LLMs, we next sought to understand the functional importance of these learned concepts.

\textbf{RQ3: How do different comment concepts affect LLM's performance?}

We conducted experiments on three SE tasks: code translation \cite{pan2024lost}, code completion with cross-file context \cite{NEURIPS2023_920f2dce}, and code refinement \cite{guo2024code-refinement}. Our objective was to assess how the activation and deactivation of individual comment concepts influence LLM performance. To this end, we first evaluated each LLM’s performance on commented records under two conditions: once with comments present and once with comments removed. Then, for records containing comments, we selectively deactivated each comment concept (e.g., Javadocs), and for records without comments, we activated each concept individually. This design allowed us to isolate the effect of each comment type on model behavior across tasks.
Our results showed that both the general concept of comments and specific comment types affect LLM performance differently across tasks and models. However, because each task was evaluated on a different set of code inputs, we could not attribute the observed differences solely to the nature of the task. To isolate the effect of the task itself, it was necessary to control for input by applying different SE tasks to the same code samples. Accordingly, we designed our final research question to investigate whether the activation of comment concepts in LLMs is conditioned on the task being performed.

\textbf{RQ4: Is the activation of different comment concepts dependent on the task?}




Such an analysis reveals task-specific conceptual dependencies on comments, which has direct implications for automated comment augmentation. Specifically, by identifying which types of comments are most impactful for a given task, augmentation systems can selectively generate or inject only the most relevant comment types-enhancing model performance without bloating input prompts. These insights can also guide developers toward writing more targeted documentation, focusing comments where they most benefit LLM-assisted tools.

The remainder of this paper is structured as follows. In \hyperref[related-work]{Section II}, we review related research relevant to our study. \hyperref[method]{Section III} describes our methodology for addressing all research questions. \hyperref[results]{Section IV} presents the results of our experiments. \hyperref[implications]{Section V} discusses the implications of our findings for both researchers and practitioners. In \hyperref[threats]{Section VI}, we outline potential threats to the validity and describe the steps taken to mitigate them. Finally, \hyperref[conclusion]{Section VII} concludes the paper and suggests directions for future work.

\section{Related Work} \label{related-work}

\subsection{Role of Comments in LLM Performance}

Prior work has examined how the presence or absence of comments affects LLM performance during inference. Nikiema et al. \cite{nikiema2025codebarrierllmsactually} investigated LLMs' sensitivity to semantic cues by applying controlled obfuscations to source code. They assessed model understanding through code description generation and deobfuscation tasks, finding that LLMs strongly rely on semantically meaningful code elements such as descriptive variable names and unobfuscated literals. However, comment inclusion showed no statistically significant effect on performance in their benchmarks. Macke and Doyle \cite{macke2024testing} used unit test generation to probe LLMs' code understanding and found that incorrect comments can substantially degrade model performance, whereas missing or incomplete comments had relatively minor impact. Similarly, Hossain et al. \cite{hossain2025tog} studied test oracle generation and demonstrated that prompts containing only Javadoc comments can match or exceed the performance of those with richer code context. Liu et al. \cite{liu2024repoqaevaluatinglongcontext} introduced the RepoQA benchmark to evaluate long-context code understanding. Their ablation study revealed that, for most LLMs, removing comments improved performance-an effect not observed in Gemini models. Likewise, Imani et al. \cite{imani2025omega} found that excluding comments enhanced the accuracy of LLM-generated code summaries.

\subsection{Pretraining to Improve LLM Performance}
A second line of research examines how incorporating comments during pretraining can enhance LLM performance. Song et al. \cite{song-etal-2024-code} showed that increasing the density of comments in the training corpus leads to improved model performance across downstream tasks. Zuo et al. \cite{zuo2025masking} proposed Masked Comment Prediction to evaluate the contribution of various code token types, finding that comments are among the most beneficial for enhancing CodeBERT’s capabilities.
Pei et al. \cite{pei2022contrastive} introduced C3P, a contrastive code-comment pretraining framework that encodes code and comments using separate encoders and aligns them in a shared multimodal embedding space to learn richer joint representations. Similarly, Chen et al. \cite{chen2024commentsnaturallogicpivots} developed MANGO, which combines contrastive comment training with a logical decoding strategy, demonstrating significant performance gains for small LLMs.

While prior studies provide useful evidence on how comment inclusion influences downstream performance through input-level augmentation and specialized training objectives, they typically treat comments as surface-level input features and focus on observable output differences. Such approaches can conflate the effects of comment presence with their semantics or quality, making it difficult to disentangle whether models rely on the concept of comments, independent of content. Moreover, output-based evaluation alone may fail to capture how LLMs internalize and generalize from these inputs. Our work fills this gap by examining whether comments-and specific types of comments-are internalized as abstract concepts, and whether activating or perturbing these concepts affects LLM behavior.

Recent advances in LLM interpretability have increasingly focused on analyzing internal representations rather than surface behavior \cite{gao2025evaluatebiasmanualtest}. Among these, circuit tracing \cite{ameisen2025circuit} has emerged as a promising technique but remains computationally prohibitive for our study, as discussed in the Introduction. An alternative and more scalable approach centers on concept-level representations. A foundational method in this space is the Concept Activation Vector (CAV) introduced by Kim et al.\cite{kim2018cav}, which tests whether a model’s latent space encodes a human-interpretable concept. The core idea is that if a linear classifier can reliably separate internal representations-i.e., the hidden-layer outputs generated as the model processes input tokens-corresponding to the presence versus absence of a concept (for instance, distinguishing inputs containing a comment from those without), then the vector orthogonal to the decision boundary (the CAV) can be utilized as a proxy to measure the internalization of the concept in model’s latent space. CAVs have since been extended to LLMs to uncover social biases \cite{gao2025evaluatebiasmanualtest}, steer model outputs \cite{turner2024steeringlanguagemodelsactivation, zhang2025controllinglargelanguagemodels}, and diagnose safety risks \cite{xu2024safetyrisks}.


Building on these advances, we adopt a representation-level perspective to investigate the conceptual role of comments in LLMs. Specifically, we examine how comments are encoded in the latent space and test whether models rely on comment concepts beyond their surface tokens. Using embedding-space perturbations, we simulate the presence or absence of comments and assess whether these internal manipulations yield behavioral effects similar to explicit input changes. This approach enables a more precise analysis of comment sensitivity and opens new pathways for steering model behavior without modifying prompts.

\section{Methodology} \label{method}
\begin{figure*}[t!]
  \centering
  \includegraphics[width=1\textwidth]{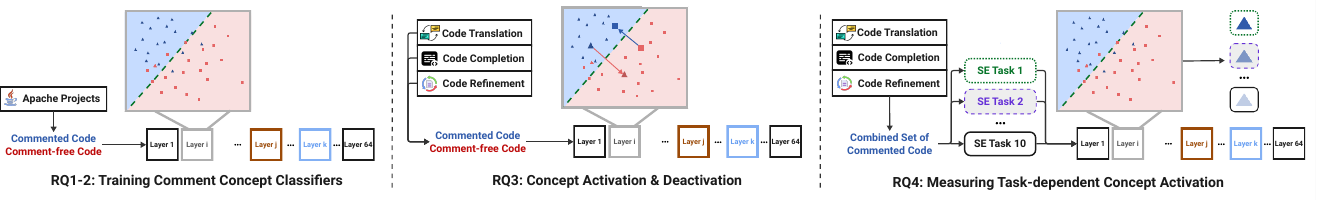}
  \caption{Overview of our Methodology. \includegraphics[width=1em]{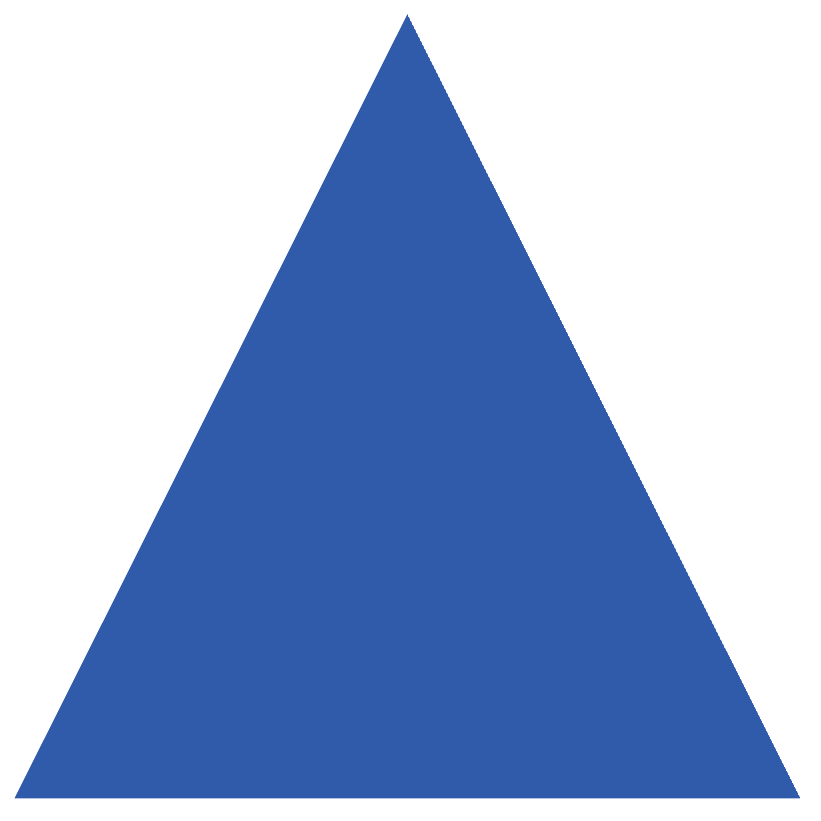} denotes embedding vectors that contain the concept $C$ and its different shades encodes its level of activation of $C$.}
  \label{fig:overall}
\end{figure*}

We begin by outlining the experimental setup used throughout our analyses, and then describe 
our methodology, as illustrated in Figure \ref{fig:overall}.


\subsection{Experimental Setup}

\textbf{Selected LLMs.} 
Prior research has demonstrated that LLM behavior can differ markedly depending on post-training objectives and data, even when the underlying architecture remains the same \cite{qwen2025qwen25technicalreport, hui2024qwen25codertechnicalreport}. For instance, instruction-tuned models are optimized to follow natural language prompts \cite{wu-etal-2024-language}, code-focused models are specialized for syntax-sensitive tasks, and reasoning-oriented models are trained to support multi-step inference through techniques that encourage structured reasoning \cite{li202512surveyreasoning}. These differences can influence how models attend to specific input structures, such as code comments. To account for this variation, we conducted our analyses across three representative model variants:  a generic Instruct LLM, a code LLM, and a reasoning LLM.



Our analyses required access to the internal representations of LLMs, necessitating the use of open-source models. To identify a high-performing LLM family suitable for our study, we referred to BigCodeBench, a comprehensive benchmark spanning 1,140 fine-grained tasks involving tool use across 139 libraries and 7 domains \cite{zhuo2025bigcodebenchbenchmarkingcodegeneration}. Our selection criteria prioritized top-tier models that belong to families encompassing all three desired variants: a generic instruction-tuned model, a code-specialized model, and a reasoning-oriented model. At the time of selection, Qwen2.5-Coder-32B-Instruct \cite{hui2024qwen25codertechnicalreport} ranked fourth overall on BigCodeBench \cite{bigcodebench-11-24}, making it a strong candidate. Importantly, it belongs to the Qwen2.5 model family \cite{qwen2025qwen25technicalreport}, which also includes the generic instruction-tuned variant (Qwen2.5-32B-Instruct) and the reasoning-augmented variant (QwQ-32B) \cite{QwenTeamQwQ-32B:Qwen}, enabling consistent cross-variant comparisons within a unified architecture.

The server on which we planned to run the experiments was equipped with an NVIDIA RTX A6000 GPU with 48GB of memory. Since it could not run models with more than 20B parameters in full precision, we opted to use the 4-bit quantized version of each LLM. This required selecting a suitable quantization method. Among the state-of-the-art quantization methods, we chose Activation-aware Weight Quantization (AWQ) due to its minimal impact on model perplexity and the inference speedup it provides for the quantized LLM \cite{Lin2023AWQ:Acceleration}. Hence, we used the AWQ-quantized version of each of the three LLMs from HuggingFace \cite{wolf2020huggingfacestransformersstateoftheartnatural}: Qwen2.5-Coder-32B-Instruct-AWQ\cite{Qwen/Qwen2.5-Coder-32B-Instruct-AWQFace}, Qwen2.5-32B-Instruct-AWQ \cite{Qwen/Qwen2.5-32B-Instruct-AWQFace}, and QwQ-32B-AWQ \cite{Qwen/QwQ-32B-AWQFace}. For simplicity, we refer to these models as the Code LLM, the Generic LLM, and the Reasoning LLM, respectively.

\textbf{LLM Deployment \& Inference.} We ran all models using SGLang \cite{zheng2204sglang} due to its state-of-the-art throughput. 
For inference hyperparameters, we followed the recommended settings for the Reasoning LLM (temperature = 0.6 and top\_p = 0.95) to avoid endless repetitions \cite{Qwen/QwQ-32B-AWQFace}. Since no recommended settings were available for the other two LLMs, we followed prior work and set the temperature to 0 to ensure deterministic output and reproducible results \cite{imani2025omega, Ouyang_2025}.
Alternatively, we could have used a non-zero temperature (e.g., 0.5) and reported the average performance across multiple stochastic runs (e.g., ten rounds at temperature 0.5). However, this approach would have exponentially increased the number of experiments in RQ3 from 36 experiments per run (3 LLMs × 3 tasks × 4 settings) to 360 experiments, making it infeasible within our resource and time constraints. Therefore, following established best practices in the field \cite{imani2025omega, Ouyang_2025}, we set the temperature to 0 to ensure result stability and reproducibility across single-run evaluations.

\textbf{Tasks.} 
To evaluate the functional role of comment concepts, we required SE tasks where the inputs consist of source code. We selected tasks that their dataset satisfies the following criteria: (1) the code must be in Java, one of the most widely used programming languages; (2) the dataset must include instances containing comments; and (3) it must provide well-defined ground truth outputs. We further prioritized SE tasks for which state-of-the-art LLM-based approaches already exist, enabling us to adopt their prompt formats and evaluation protocols.

Based on these criteria, we selected three distinct tasks: code translation \cite{pan2024lost}, code refinement \cite{guo2024code-refinement}, and code completion \cite{cceval}. The code translation task uses a dataset consisting of Java programs, test inputs, and corresponding expected outputs \cite{pan2024lost}. It supports translation into multiple programming languages. In our study, we selected four target languages-C, C++, Python, and Go-based on their popularity in software development \cite{TIOBETIOBE}.
The code refinement task \cite{guo2024code-refinement} presents Java code snippets with accompanying review comments, testing the model's ability to revise code in response to feedback. The code completion \cite{cceval} task assesses repository-level completion by providing partial Java code snippets along with relevant context from surrounding files within the same repository, simulating realistic multi-file development settings.

To identify records containing comments and to classify comment types (i.e., Javadoc, Inline, and Multiline), we employed Nirjas \cite{nirjas}, a language-agnostic tool for comment analysis that supports Java. As part of our experimental design, we compared original inputs with their comment-free counterparts. To ensure meaningful comparisons, we excluded records from the code refinement task where the only modifications to the original code were within comments. For the code completion benchmark, we retained the instances where either the partial code snippet or the retrieved context contained at least one comment. From the filtered set of commented records in each benchmark, we randomly sampled instances such that the sample size satisfied statistical power requirements (95\% confidence level, 5\% margin of error). Table \ref{table:benchmarks} summarizes the post-filtering statistics for each task.



\textbf{Evaluation Metrics.} 
To ensure we accurately evaluated the performance in each task, we adopted all the metrics utilized for each task from the original work.
For the code translation task \cite{pan2024lost}, we adopted the Success Rate (SR), calculated as the ratio of correctly translated outputs (i.e., those that produce the expected output given the input) to the total number of translations. 
Regarding code refinement, we used Exact Match (EM) to compare the generated code with the expected refined version. We also employed EM-trim, a more lenient metric that considers the output correct if the beginning or end of the generated code matches the ground truth after trimming irrelevant content, and used by the state-of-the-art work \cite{guo2024code-refinement}. Additionally, we used BLEU-4 \cite{bleu}, a widely adopted metric in neural translation tasks \cite{imani2025omega, li2024omg}, which measures the overlap of 4-grams between the generated code and the ground truth. We further adopted BLEU-trim \cite{guo2024code-refinement}, which applies BLEU-4 to the trimmed version of the generated output.

For the code completion benchmark, we used four metrics introduced by the authors of the dataset \cite{cceval}, grouped into two categories: Code Match and ID Match. Code Match assesses the similarity between the generated and reference code using exact match (EM) and edit similarity (ES). ID Match focuses on the model’s ability to predict the correct application programming interfaces (APIs) by comparing the ordered list of identifiers extracted from the generated and reference completions. In this setting, EM and F1 scores are computed based on identifier overlap. The F1 score is:
$\mathbf{F_1} = \frac{\textbf{2} \cdot \textbf{TP}}{\textbf{2} \cdot \textbf{TP} + \textbf{FP} + \textbf{FN}}$
where TP is the number of correctly predicted identifiers (true positives), FP is the number of identifiers predicted by the model but not present in the reference (false positives), and FN is the number of reference identifiers missed by the model (false negatives).

To compare performance differences caused by activating or deactivating comment concepts, we used the relative delta metric. Absolute score changes can be misleading, especially across tasks or metrics with different scales or baselines, and fail to reflect meaningful shifts when original values are small \cite{mayrand1996experiment}. For example, a change from 10 to 12 represents a 20\% improvement, while a change from 90 to 92 is only about 2\%, despite having the same absolute difference. Relative delta normalizes the change with respect to the original performance, capturing the proportional impact and enabling fairer comparisons across tasks, models, and comment types.
\[\Delta_{rel}=\frac{P_{modified} - P_{original}}{P_{original}} \times 100\]
where $P_{original}$ is the performance before treatment and \\$P_{modified}$ is the performance after a treatment.

\begin{table}[t!]
\caption{Statistics of the Adopted Benchmarks}
\label{table:benchmarks}
\centering
\resizebox{\columnwidth}{!}{%
\begin{tabular}{lcccc}
\toprule
\textbf{Benchmark} & \textbf{Sample Size} & \textbf{\#Javadocs}  & \textbf{\#Inline} & \textbf{Multiline} \\
(\#Commented/Total) & & & & \\
\midrule
Code Completion (1046/2139) & 281 & 227 & 270 & 97 \\
Code Refinement (103/14945) & 81 & 0 & 66 & 17 \\
Code Translation (47/200) & 43 & 12 & 15 & 16 \\
\bottomrule
\end{tabular}
}
\end{table}

\subsection*{RQ1-2: Analyzing Internalization of Comments}

As the first step to investigate the extent to which the LLMs are internally aware of comments and its different forms as distinct concepts, we defined four concepts to answer RQ1 and RQ2. Specifically, to answer RQ1, we defined \textcircled{\small{1}} \textit{Comment} concept as comments of any form in Java that can either start start with $/*$, end with $*/$, and span multiple lines, or single or multiple lines of comments that being with $//$.

In order to analyze the internalization of different forms of comments (RQ2), we defined separate concepts for each comment type:
\textcircled{\small{2}} \textit{Javadocs}: Comments that start with $/*$, end with $*/$, and span multiple lines. \textcircled{\small{3}} \textit{Inline Comments}: Comments that appear in code as a single line of comment. This type of comment can either start with $//$, or start with $/*$ and end with $*/$ in the same line. An inline comment can be standalone or can be at the end of a line of code. 
\textcircled{\small{4}} \textit{Multiline Comments}: Consecutive lines of inline comments. 

In order to investigate the extent to which the LLMs have internalized each concept, we adopt the interpretation approach utilized by Xu et al \cite{xu2024safetyrisks}. 
We follow the linear interpretability assumption that is widely used in the interpretability studies \cite{zhang-etal-2024-distillation, NIPS2017_dc6a7e65, Bau_2017_CVPR}. This assumption suggests that a concept $C$ is internalized by a deep model if a linear classifier, trained on embedding vectors of positive examples (containing $C$) and negative examples (without $C$), can distinguish between them with high accuracy \cite{kim2018cav, xu2024safetyrisks}. Following the approach of Xu et al. \cite{xu2024safetyrisks}, we adopted the \texttt{LogisticRegression} classifier from the scikit-learn library \cite{varoquaux2015scikit} using default parameters. 

\textbf{Classifier's Scope.} In our study, the deep model under investigation is an LLM, which consists of several hidden layers. Each of these layers can encode a variety of learned concepts \cite{zhang2023widerdeeperllmnetworks}. The LLMs we adopted in this study contain 64 hidden layers, and the comment concept, along with its subtypes, may be learned at any of them. Therefore, a linear classifier must be trained at each layer to identify where within the model's internal each concept is internalized, and how accurately each layer distinguishes these concepts.

\textbf{Classifier's Input.} A separate classifier is trained at each layer of the LLM, requiring a single representative vector derived from that layer's processing of the input code. Each layer of the LLM outputs a vector for every token in the input sequence; this contextualized output vector is known as a \textit{hidden state}. However, since a given layer produces a sequence of hidden states - one for each token in the input code - an aggregation method is required to transform these into a single representative vector. Our adopted LLMs employ a decoder-only Transformer architecture, which is constrained by a causal attention mechanism \cite{vaswani2017attention}. This mechanism ensures that a token's representation is only influenced by itself and the tokens that precede it. As a result, the hidden state of the last token is the only representation at any given layer that has been informed by the entire input sequence. Therefore, we adopt the common practice of using the hidden state of the last token \cite{behnamghader2024llm2veclargelanguagemodels, wang2023improving} as the input vector for the classifier at that layer.

\textbf{Dataset.} To prevent the linear classifier at each layer from overfitting to the datasets of adopted SE tasks \cite{ying2019overview}, we curated a separate dataset specifically for training and evaluation. Following common practice in Java-focused studies \cite{ren2023api, li2023commit, chen2017characterizing}, we sourced our data from Apache repositories. We cloned the top 10 most-starred Apache repositories and retained only their Java files. These files were processed using the same Nirjas-based \cite{nirjas} comment processing script that we applied to the benchmarks. For each Java file containing one of our defined concepts, we created two versions: one with the concept present (positive example) and one with the concept removed (negative example). For example, if a file contained Javadocs, we generated both the original version and a version with all Javadocs stripped. This resulted in 69,610 records with comments, 25,564 records with inline comments, 9,327 records with multiline comments, and 34,476 records with Javadocs.

\textbf{Train-Test Split.} Given the example set for each concept, to better understand how accurate the conecpt is encoded at each layer, we trained the linear classifier multiple times by having a fix test set and varying train size. Specifically, for each concept, we randomly sampled $S=2N$ records (750 $\pm$ 14), where $N$ is a statistically significant sample size (confidence level 95\%, margin of error 5\%). For each record, we included both the positive and negative examples in the sample. From this sample, we fixed the test set to consist of $N$ records and gradually increased the training set size from $0.01S$ up to $0.5S = N$.  This variation in train size allows us to assess how well each concept is internalized across the LLM’s layers. Achieving a high classification accuracy with a small training size indicates that the concept is strongly encoded in the model’s representations.

\subsection*{RQ3: Measuring the Importance of Concepts}
Assuming that the \textit{Comment} concept and its subtypes are internalized within the layers of LLMs, it becomes essential to investigate whether this internal representation influences model performance. Specifically, we must understand whether the presence of these conceptual encodings enhances or hinders LLMs' effectiveness in performing SE tasks. 
To investigate the functional importance
of the concepts, we first ran each task - code translation, code completion, and code refinement - on samples corresponding to each comment type. For each sample, we executed the benchmark twice: once on the original code ($e1$) and once on the same code with the specific comment type removed ($e2$) (e.g., Javadocs for the Javadoc sample). This allowed us to establish the baseline performance of the LLM in the presence and absence of each concept in the input. 

Next, we aimed to design experiments to measure how the activation or deactivation of different comment concepts influences model behavior. One possible approach was to use prompt-level instructions, such as asking the LLM to ignore comments for $e1$ or inject additional comments for $e2$ while performing the software engineering task. However, although LLMs have improved in instruction following, their compliance with prompt directives remains inconsistent \cite{heo2025do}. To avoid this inconsistency and isolate the effect of internalized concepts, we instead followed prior work \cite{xu2024safetyrisks, heo2025do} and employed an embedding-level intervention.

Given $e1$, if we can manipulate the internal representation such that the LLM interprets it as lacking a concept (e.g., the \textit{Comment} concept), without removing the concept from the input, we can isolate and reveal the role of the internalized concept in driving the LLM's behavior. We refer to this experiment as Concept Deactivation (CD).
Conversely, starting with $e2$, an embedding derived from an input that does not contain the concept, if we shift its internal representation so that the LLM interprets it as if the concept were present, we can examine whether the concept’s influence on performance stems solely from its internal encoding rather than its presence in the input. We name this procedure Concept Activation (CA).
Together, two complementary experiments help disentangle input-level and representation-level effects. In the following, we explain how CD and CA were made possible using the linear classifiers we trained for each concept at each layer of the LLMs.

\begin{figure}[t!]
  \centering
  \includegraphics[width=0.62\linewidth]{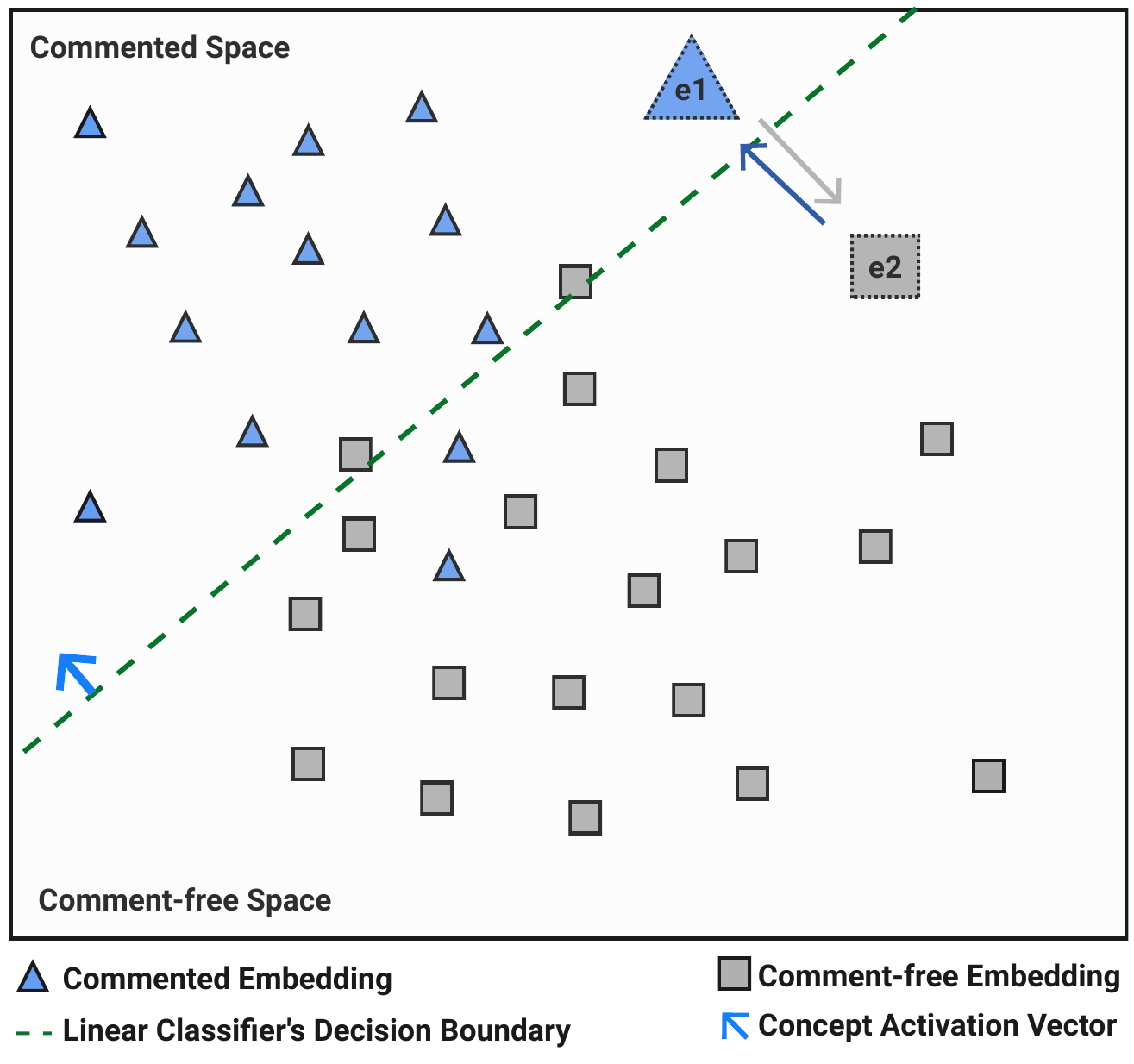}
  \caption{Activation/Deactivation of Comment Concepts}
  \label{fig:cav}
\end{figure}

Considering a linear classifier $h_l(C)$ that is trained at the $l_{th}$ layer of an LLM to separate inputs containing concept $C$, e.g. the Comment concept, from those that lack $C$ with a high accuracy, its decision boundary can reliably separate inputs that contain a concept $C$ from those that do not. To illustrate this, Figure~\ref{fig:cav} visualizes the embedding space of a hypothetical LLM layer with a 2D representation. Given a trained classifier that separates embeddings of commented and comment-free inputs, we can identify a vector orthogonal to the decision boundary that points toward the region where embeddings with the concept reside. This vector is referred to as the Concept Activation Vector (CAV). 

Recalling $e2$, the representation of a comment-free code at the $l_{th}$ layer, we can shift it along the direction of the CAV with sufficient magnitude to move it into the region associated with commented inputs. By doing so, we make the LLM internally interpret $e2$ as if it contains the \textit{Comment} concept, i.e., we \textit{activate} the comment concept for $e2$. An opposite shift can be applied to $e1$, the representation of a commented code at the $l_{th}$ layer, by moving $e1$ against the direction of the CAV, thereby \textit{deactivating} the \textit{Comment} concept in $e1$. 
Throughout applying this layer-wise embedding perturbation guided by the CAV for each concept at all layers where the classifier $h_l(C)$ achieved reliable accuracy, we performed the previously described Concept Deactivation (CD) and Concept Activation (CA) experiments. These experiments were designed to assess the functional importance of the internalized concept on the LLMs’ performance. We formally present the algorithm below.

Given the classifier at layer $l$ of an LLM for concept $C$, the activeness of the concept $C$ at $l$ for an embedding $e$ can be measured as $\mathbf{P_c(e)}$, which is the prediction probability of the linear classifier:
\[ \mathbf{P_c(e) = Sigmoid(w^T \cdot e+b)} \]
where $\mathbf{w} \in \mathbb{R}$, $\mathbf{b} \in \mathbb{R}$, are the learned parameters of the classifier. 

Algorithm \ref{alg:perturbation} presents our approach to activate and deactivate comment concepts.
Given the value of $\mathbf{P_c(e)}$ for a given embedding $\mathbf{e}$, we apply perturbations at layers where the concept classifier’s accuracy meets or exceeds a threshold $\mathbf{T}$.
This threshold ensures that perturbations are only applied in layers where the CAV reliably captures the direction of the concept in the embedding space, making the activation or deactivation meaningful and trustworthy.
To deactivate the concept ($D=\text{against}$), we shift the embedding in the opposite direction of the concept's CAV such that the new predicted probability $\mathbf{P_{e'}}$ drops to a target probability $\mathbf{P_t}$  ($\mathbf{P_{e'}} \leq \mathbf{P_t}$). Conversely, to activate the concept, we shift the embedding toward the direction of CAV until $\mathbf{P_{e^\prime}}$ increases to $\mathbf{P_t}$  ($\mathbf{P_{e'}} \geq \mathbf{P_t}$). 

In both activation and deactivation settings, we perturbed the original embedding $\mathbf{e}$ by adding a scaled CAV: $\mathbf{e}' = \mathbf{e} + \epsilon \cdot \mathbf{v}$. The unit vector $\mathbf{v}$ ($|\mathbf{v}| = 1$) defines the direction in embedding space that increases or decreases the classifier’s predicted probability $P_c(\mathbf{e}')$ toward a target value $P_t$. To minimize interference with the model’s learned representations, the scalar $\epsilon$ is computed by solving a constrained optimization problem, selecting the smallest perturbation necessary to achieve $P_c(\mathbf{e}') = P_t$ with minimal impact on downstream model loss, as proposed by Xu et al. \cite{xu2024safetyrisks}.
We note that while introducing random perturbations in arbitrary directions within the embedding space could serve as a baseline to compare against our guided perturbations, recent studies in established NLP venues \cite{xu2024safetyrisks, heo2025do} have shown that CAV-based guided perturbations reliably steer model representations toward or away from specific concepts (e.g., safety). 
Given the compelling evidence for the effectiveness of guided perturbations in prior work \cite{xu2024safetyrisks, heo2025do}, employing random perturbations as baselines is unnecessary.

To ensure a fair comparison across all comment concepts and LLMs in the CD and CA experiments, it is important that the selection of activated layers is not biased by disparities in concept classification accuracy across LLMs and concepts. Using a fixed threshold, such as $T = 0.90$, could result in an uneven number of qualifying layers for different concepts or LLMs, skewing the effects of activation or deactivation. To address this, we adopted a dynamic approach: we set the threshold $T$ to the minimum of the medians of classifier accuracies observed across all comment concepts and models, which was 0.84 (See \hyperref[results]{Section IV} for details about classification accuracies). For the target probability $P_t$, we followed the setting used by Xu et al. \cite{xu2024safetyrisks}, adopting $P_t = 0.01$ for concept deactivation to ensure minimal activation and $P_t = 1 - 0.01 = 0.99$ for a concept activation.


\begin{algorithm}[t!]
\caption{Conceptual Activation/Deactivation of Comments}\label{alg:perturbation}
\begin{algorithmic}
\Require Input code $x$, LLM with $L$ layers, Direction $D$, Classifier $C_l$, Accuracy Threshold $T$, Target Probability $P_t$
\For{$l=1$ to $L$}
\If{$TestAcc(C_l)>T$}
    \State $e \gets$ Embedding of $x$ at the $l_{th}$ layer \textit{after} activating or deactivating the previous layers
    \If{($P_c(e)>P_t$ \& $D=against$) || ($P_c(e)<P_t$ \& $D=toward$)}
        \State $e^\prime \gets e + \epsilon \cdot v$
    \EndIf
\EndIf
\EndFor
\end{algorithmic}
\end{algorithm}

\subsection*{RQ4: Role of Task on the Activation of Concepts}

Our experiments in RQ3 reveals the functional role of the activeness of different comment concepts in shaping LLMs’ performance  across three SE tasks, each involving different code inputs. Given this impact, it becomes essential to investigate whether and how the requested SE task itself, independent from the code, influences the activation of these concepts within each LLM. Understanding this relationship allows us to examine the extent to which internalized concepts are sensitive to task, shedding light on potential task-concept dependencies and how LLMs adjust their internal representations in response to the requested task. 

Therefore, for each comment concept, we merged the corresponding sample from all three tasks to form a fixed set of code snippets $\mathcal{C} = {c_1, c_2, \dots, c_n}$. To promote generalization while maintaining feasibility, we selected a representative subset of 10 tasks rather than exhaustively covering all possible SE tasks. We grounded our selection in the categorization proposed by Hu et al.\cite{hu2025assessing}, who surveyed 191 SE benchmarks. Specifically, we reviewed the domains identified in their taxonomy and extracted tasks where source code serves as the primary input. We collected the set of tasks $\mathcal{T} = {t_1, t_2, \dots, t_{10}}$ that are listed as: \textcircled{\small{1}} Code Summarization \textcircled{\small{2}} Code Translation \textcircled{\small{3}} Test Generation \textcircled{\small{4}} Code Completion \textcircled{\small{5}} Fault Localization \textcircled{\small{6}} Program Repair \textcircled{\small{7}} Vulnerability Detection \textcircled{\small{8}} Code Review \textcircled{\small{9}} Code Refactoring \textcircled{\small{10}} Code Documentation.  


Recent studies on prompt engineering for SE tasks have shown that zero-shot prompts are sufficient to capture the semantics of a task and that surface-level variations in prompt phrasing have limited impact on model behavior \cite{wang2024advancedlanguagemodelseliminate, imani2025omega}. Therefore, for each task, we designed a minimal zero-shot task instruction that succinctly described the task without introducing guidance or unnecessary detail. This approach helps ensure that any observed differences in concept activation stem from the task itself rather than prompt formulation. For example, for Program Repair, we used the prompt ``\textit{Fix the bug in the following code snippet to make it work as intended.}''. We have included the prompts we used for all tasks in supplementary \cite{SupplementaryMaterials}. Given the set of task instructions $\mathcal{T}$ and the fixed set of code inputs $\mathcal{C}$, we formed a prompt set $\mathcal{P}$ for each task where $\mathcal{P}_j = {(t_j, c_i) \mid c_i \in \mathcal{C},\ t_j \in \mathcal{T} }$, ensuring that every code snippet was processed under each task formulation.

Given the prompt set $\mathcal{P}_j$ for each of the 10 tasks, we inputted each prompt into the LLM and captured its representation $e_i$ at every layer $l$, following the same procedure used to generate inputs for the concept classifiers. Using the layer-specific concept classifier $h_l$, we then computed the concept activation value for each $e_i$ as the probability that $h_l$ assigns $e_i$ to the embedding region corresponding to the presence of the concept ($P_c(e)$). Finally, we averaged these activation values across all prompts in $\mathcal{P}_j$ to obtain the mean activation for each task, and compared these values across LLMs for each comment concept.

\begin{figure*}[t!]
\Description[Test Accuracy of Comment Concept Classifiers at Each Layer of the Three LLMs]{}
  \centering
  \includegraphics[width=1\textwidth]{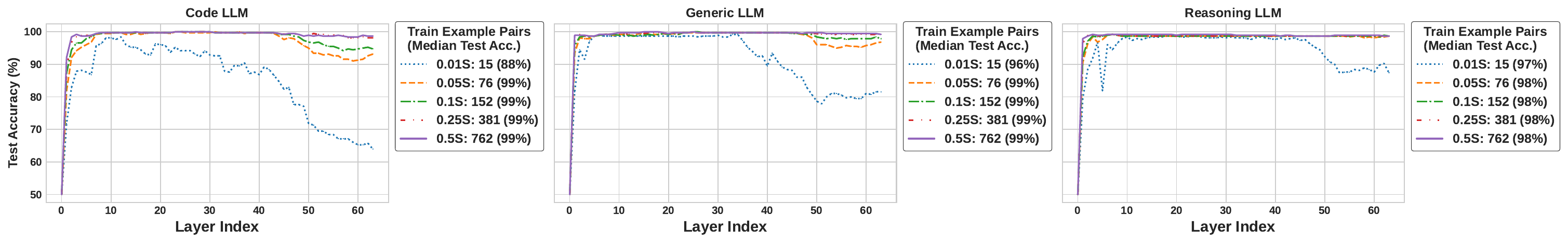}
  \caption{Test Accuracy of Comment Concept Classifiers at Each Layer of the Three LLMs
}
\label{fig:r1-results}
\end{figure*}
\begin{figure*}[b!]
\Description[Test Accuracy of Javadoc, Inline, and Multiline Concept Classifiers at Each Layer of the Three LLMs]{}
  \centering
  \includegraphics[width=1.0\textwidth]{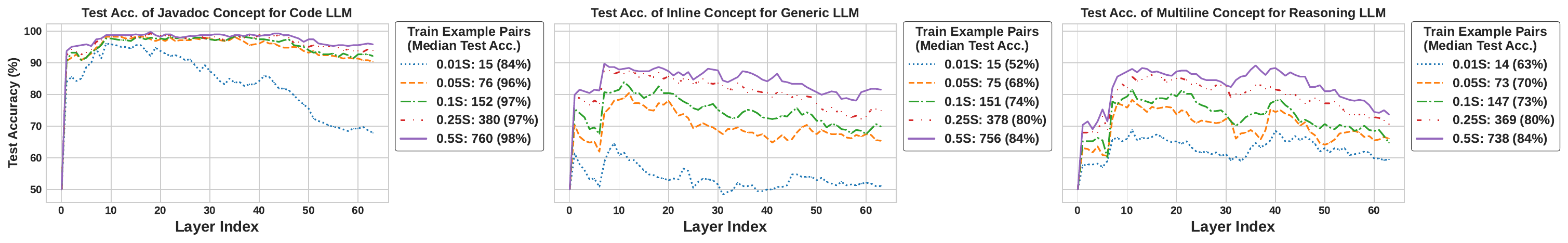}
  \caption{Test Accuracy of Javadoc, Inline, and Multiline Concept Classifiers at Each Layer of the Three LLMs
}
\label{fig:r2-results}
\end{figure*}

\section{Results} \label{results}
In this section, we iterate our research questions and answer them by providing the results of experiments presented in \hyperref[method]{Section III}.

\textbf{RQ1: Do LLMs internally recognize comments as a concept?}

As described in \hyperref[method]{Section III}, we trained a linear classifier at each layer of the LLMs to distinguish between examples with and without comments, using varying training set sizes. This allowed us to quantitatively assess the extent to which comments are encoded as a separable concept in the models’ internal representations.

Figure \ref{fig:r1-results} presents the classification accuracy of the comment concept classifiers across all layers of each LLM. Remarkably, all three LLMs achieved a median test accuracy above 90\% with just 5\% of the full training set (15 example pairs with and without comments, as defined in \hyperref[method]{Section III}. Notably, the Reasoning LLM attains the highest median accuracy of 97\% even at the smallest training size ($0.01S$), indicating that comment information is strongly encoded even with minimal supervision.


\begin{boxH}
\textbf{RQ1 Answer:} 
LLMs internally encode comments as a distinct and robustly separable concept.
\end{boxH}

\textbf{RQ2: Do LLMs internalize different types of comments as distinct concepts?}

Similar to RQ1, we utilize the accuracy of classifiers trained on example sets from each comment type to examine whether LLMs internalize comment types (i.e., inline, multi-line, and Javadocs) as distinct concepts. Due to space constraints and the similarity in classification trends across the LLMs, we present the accuracy results for each concept using only one representative LLM. Full classification accuracy results for all models are provided in the supplementary material \cite{SupplementaryMaterials}.

Figure \ref{fig:r2-results} shows the median classification accuracy of three distinct comment types, namely, Javadoc, Inline, and Multiline, across all LLM layers. Consistent with the overall comment concept results, Javadocs are strongly internalized by the Code LLM, achieving a median classification accuracy of 96\% with 76 pairs of examples. In comparison, the Generic and Reasoning LLMs exhibit similar accuracy trends for Inline and Multiline comments. Notably, under the smallest training size, Multiline comments are more distinguishable than Inline comments, as reflected in the higher median classification accuracy of 63\% compared to 52\% (see the middle plot, lowest dotted line). However, even when using the entire training set, the median accuracy for Inline and Multiline comments (both 84\%) only approaches the level achieved by Javadocs with only 15 example pairs. These results indicate that Javadocs are more robustly encoded as a distinct concept in LLMs, particularly in code-specialized LLM, suggesting a greater model sensitivity to semantically rich documentation compared to simpler comment forms.



\begin{boxH}
\textbf{RQ2 Answer:} LLMs internally recognize different comment types as distinct concepts, with Javadocs as the most robustly internalized concept.
\end{boxH}

\textbf{RQ3: How do different comment concepts affect LLM's performance?}

To answer this RQ, as detailed in \hyperref[method]{Section III}, we conducted two experiments for each of our adopted benchmarks—code translation, code completion, and code refinement. In the first experiment, \textbf{C}oncept \textbf{D}eactivation (CD), we evaluated how LLM performance changes when individual comment concepts are deactivated in code that originally includes comments, compared to when those concepts remain active. 
In the second experiment, \textbf{C}oncept \textbf{A}ctivation (CA), we focused on comment-free records and assessed the effect of activating each comment concept in the absence of explicit comments. Below, we present the results of these two experiments across all benchmarks.

\begin{table*}[t!]
\caption{RQ3 results for Code Completion. Bold: Largest impact of activating or deactivating each concept. \highlighted{Cell}: Largest overall impact on each experiment and metric.}
\label{tab: rq3-completion}
\centering
\resizebox{0.8\textwidth}{!}{
\begin{tabular}{llll|ll}
\toprule
\multirow{2}{*}{\textbf{Concept}} &\multirow{2}{*}{\textbf{LLM}} &\multicolumn{2}{c|}{Commented EM / CD EM ($\Delta_{rel}$)} &\multicolumn{2}{c}{Comment-free EM / CA EM  ($\Delta_{rel}$)} \\
& &\textbf{Code Match} &\textbf{ID Match} &\textbf{Code Match} &\textbf{ID Match} \\\midrule
\multirow{3}{*}{Comment} &Code LLM &13.88 / 13.17 (-5.12\%) &32.38 / 30.96 (-4.39) &\textbf{12.81 /10.32 (-19.44\%)} &\textbf{30.96 / 28.83 (-6.88\%)} \\
&Generic LLM &18.15 / 18.15 (0.00\%) &31.32 / 30.60 (-2.30\%) &16.01 / 17.08 (6.68\%) &29.18 / 28.11 (-3.67\%) \\
&Reasoning LLM &\textbf{13.88 / 13.17 (-5.12\%)} &\textbf{27.76 / 24.56 (-11.53\%)} &11.39 / 11.74 (3.07\%) &23.49 / 23.49 (0.00\%) \\
\midrule
\multirow{3}{*}{Javadoc} &Code LLM &12.33 / 12.33 (0.00\%) &30.4 / 28.63 (-5.82\%) &\textbf{12.81 / 10.32 (-19.44\%)} &\textbf{30.96 / 29.18 (-5.75\%)} \\
&Generic LLM &16.3 / 16.3 (0.00\%) &29.52 / 28.63 (-3.01\%) &16.01 / 17.44 (8.93\%) &29.18 / 28.11 (-3.67\%) \\
&Reasoning LLM &\textbf{12.78 / 11.45 (-10.41\%)} &\textbf{27.31 / 24.67 (-9.67\%)} &11.39 / 11.03 (-3.16\%) &23.49 / 23.49 (0.00\%) \\
\midrule
\multirow{3}{*}{Inline} &Code LLM &\textbf{14.07 / 12.96 (-7.89\%)} &32.96 / 30.37 (-7.86\%) &\cellcolor{rowgray}\textbf{12.81 / 9.96 (-22.25\%)} &\textbf{30.96 / 28.47 (-8.04\%)} \\
&Generic LLM &18.15 / 19.26 (6.12\%) &31.48 / 31.11 (-1.18\%) &16.01 / 16.37 (2.25\%) &29.18 / 27.76 (-4.87\%) \\
&Reasoning LLM &14.44 / 13.33 (-7.69\%) &\cellcolor{rowgray}\textbf{28.52 / 24.81 (-13.01\%)} &11.39 / 11.03 (-3.16\%) &23.49 / 25.27 (7.58\%) \\
\midrule
\multirow{3}{*}{Multiline} &Code LLM &13.4 / 14.43 (7.69\%) &34.02 / 31.96 (-6.06\%) &\textbf{12.81 /10.32 (-19.44\%)} &30.96 / 28.83 (-6.88\%) \\
&Generic LLM &15.46 / 14.43 (-6.66\%) &32.99 / 31.96 (-3.12\%) &16.01 / 16.73 (4.50\%) &29.18 / 28.83 (-1.20\%) \\
&Reasoning LLM &\cellcolor{rowgray}\textbf{13.4 / 11.34 (-15.37\%)} &\textbf{26.8 / 28.87 (7.72\%)} &11.39 / 13.52 (18.70\%) &\cellcolor{rowgray}\textbf{23.49 / 27.76 (18.18\%)} \\
\bottomrule
\end{tabular}
}
\end{table*}

\begin{table}[b!]
\caption{RQ3 results for Code Translation to Python. SR: Success Rate, $\mathbf{SR^\alpha}$: Commented SR, $\mathbf{SR^\gamma}$: Comment-free SR. Bold: Largest impact of activating or deactivating a concept. \highlighted{Cell}: Largest overall impact from activation and deactivation across concepts and LLMs.}
\label{tab: rq3-translation}
\centering
\resizebox{1\columnwidth}{!}{
\begin{tabular}{llll}
\toprule
\textbf{Concept} &\textbf{LLM} &\textbf{$\mathbf{SR^\alpha}$ / CD SR ($\Delta_{rel}$)} &$\mathbf{SR^\gamma}$\textbf{ / CA SR  ($\Delta_{rel}$)} \\\midrule
\multirow{3}{*}{Comment} &Code &51.16 / 53.49 (4.55\%) &\cellcolor{rowgray}\textbf{46.51 / 60.47 (30\%)} \\
&Generic &\textbf{39.53 / 48.84 (23.53\%)} &41.86 / 51.16 (22.22\%) \\
&Reasoning &76.74 / 86.05 (12.12\%) &83.72 / 69.77 (-16.67\%) \\
\midrule
\multirow{3}{*}{Javadoc} &Code &66.67 / 58.33 (-12.5\%) &\textbf{46.51 / 51.16 (10.00\%)} \\
&Generic &\cellcolor{rowgray}\textbf{25.00 / 41.67 (66.67\%)}&41.86 / 44.19 (5.56\%) \\
&Reasoning &66.67 / 91.67 (37.5\%) &83.72 / 88.37 (5.56\%) \\
\midrule
\multirow{3}{*}{Inline} &Code &\textbf{56.25 / 68.75 (22.22\%)} &46.51 / 51.16 (10.00\%) \\
&Generic &50.00 / 56.25 (12.50\%) &\textbf{41.86 / 48.84 (16.67\%)} \\
&Reasoning &87.50 / 93.75 (7.14\%) &83.72 / 81.40 (-2.78\%) \\
\midrule
\multirow{3}{*}{Multiline} &Code &33.33 / 40.00 (20\%) &46.51 / 51.16 (10.00\%) \\
&Generic &\textbf{40.00 / 53.33 (33.33\%)} &41.86 / 44.19 (5.56\%) \\
&Reasoning &73.33 / 73.33 (0.00\%) &\textbf{83.72 / 69.77 (-16.67\%)} \\
\bottomrule
\end{tabular}
}
\end{table}

Table \ref{tab: rq3-translation} presents the results of our experiments on the code translation benchmark for translating Java to Python. Due to space limitations, results for translations to C, C++, and Go are omitted here and included in the supplementary material \cite{SupplementaryMaterials}. 


In the CD experiment, we find that deactivating comment concepts generally improves translation performance across LLMs, with the exception of Javadoc deactivation in the Code LLM, which led to a performance drop. The most substantial gain was observed in the Generic LLM, where deactivating the Javadoc concept resulted in a 66.67\% improvement. Interestingly, this contrasts with the performance decline seen in the Code LLM under the same condition. In the CA experiment, the effect of activating comment concepts varied across LLMs and concepts. The most notable improvement occurred when activating the general Comment concept in the Code LLM, yielding a 30\% performance boost.

Table \ref{tab: rq3-completion} presents changes in LLMs’ Exact Match (EM) scores for the code completion task under the CD and CA experiments. Due to space constraints, changes in Edit Similarity and F1 score are included in the supplementary material \cite{SupplementaryMaterials}. In contrast to the translation task, deactivating comment concepts in code completion typically results in a performance decline, with losses reaching up to 15\% in several cases. However, the overall magnitude of change is smaller compared to the code translation benchmark. Among the three LLMs, the Reasoning LLM experiences the largest performance drops in three out of four concept deactivations. Interestingly, while the Code LLM benefited most from concept activation in the translation task, it exhibits the most negative impact from activation in the completion task. The most substantial gain in the CA experiment is observed in the Reasoning LLM, where activating the Multiline Comment concept improves ID Match EM by 18.18\%.

The impact of CD and CA on the code refinement benchmark is presented in Table \ref{tab: rq3-code-refinement}. EM is not included in the table since no changes were observed across any condition. A consistent negative impact on BLEU score is seen when deactivating comment concepts, with the Reasoning LLM being the most severely affected across all LLMs and tasks. While activating the concepts hurt the performance of both the Code LLM and the Reasoning LLM, it improved the performance of the Generic LLM by up to 27.13\%.

Overall, deactivating comment concepts improved performance in code translation but caused a performance decline in code completion and refinement. Similarly, activating the concepts produced mixed outcomes, often enhancing translation performance while negatively impacting code completion and refinement tasks.

\begin{boxH}
\textbf{RQ3 Answer:} 
Different comment concepts impact LLMs' performance in three SE tasks from -90\% to +67\%. 
In two out of three tasks, the most substantial impact was observed for the Reasoning LLM, with Multiline comments causing the largest performance change across LLMs.
\end{boxH}

\begin{table*}[ht!]
\caption{RQ3 results for Code Refinement. Bold: Largest impact of activating or deactivating each concept. \highlighted{Cell}: Largest overall impact on each experiment and metric.}
\label{tab: rq3-code-refinement}
\centering
\resizebox{0.8\textwidth}{!}{
\begin{tabular}{llll|ll}\toprule
\multirow{2}{*}{\textbf{Concept}} &\multirow{2}{*}{\textbf{LLM}} &\multicolumn{2}{c|}{Commented Metric / CD Metric  ($\Delta_{rel}$)} &\multicolumn{2}{c}{Comment-free Metric / CA Metric  ($\Delta_{rel}$)} \\
& &\textbf{BLEU} &\textbf{BLEU\_trim} &\textbf{BLEU} &\textbf{BLEU\_trim} \\\midrule
\multirow{3}{*}{Comment} &Code LLM &70.47 / 45.90 (-34.87\%) &81.04 / 81.85 (0.99\%) &57.53 / 44.79 (-22.14\%) &\textbf{62.90 / 81.94 (30.26\%)} \\
&Generic LLM &82.76 / 66.28 (-19.92\%) &\textbf{86.69 / 81.13 (-6.42\%)} &62.93 / 68.61 (9.02\%) &64.71 / 82.26 (27.13\%) \\
&Reasoning LLM &\textbf{72.89 / 8.34 (-88.56\%)} &77.81 / 77.93 (0.16\%) &\cellcolor{rowgray}\textbf{59.82 / 7.00 (-88.30\%)} &61.05 / 55.21 (-9.57\%) \\
\midrule
\multirow{3}{*}{Inline} &Code LLM &71.25 / 44.68 (-37.28\%) &82.29 / 82.62 (0.40\%) &57.53 / 45.47 (-20.97\%) &\textbf{62.90 / 82.34 (30.91\%)} \\
&Generic LLM &83.51 / 67.88 (-18.72\%) &\textbf{86.80 / 81.63 (-5.95\%)} &62.93 / 66.95 (6.38\%) &64.71 / 80.64 (24.61\%) \\
&Reasoning LLM &\textbf{73.80 / 8.14 (-88.97\%)} &77.82 / 79.07 (1.63\%) &\textbf{59.82 / 7.05 (-88.21\%)} &61.05 / 60.56 (-0.80\%) \\
\midrule
\multirow{3}{*}{Multiline} &Code LLM &66.01 / 40.46 (-38.70\%) &76.06 / 77.26 (1.58\%) &57.53 / 44.90 (-21.96\%) &\cellcolor{rowgray}\textbf{62.90 / 82.37 (30.96\%)} \\
&Generic LLM &80.35 / 59.95 (-25.39\%) &85.71 / 75.90 (-11.45\%) &62.93 / 65.63 (4.30\%) &64.71 / 78.45 (21.24\%) \\
&Reasoning LLM &\cellcolor{rowgray}\textbf{70.57 / 6.90 (-90.23\%)} &\cellcolor{rowgray}\textbf{72.44 / 82.23 (13.51\%)} &\textbf{59.82 / 7.44 (-87.57\%)} &61.05 / 59.23 (-2.99\%) \\
\bottomrule
\end{tabular}
}
\end{table*}

\textbf{RQ4: Is the activation of different comment concepts dependent on the task?}

Table \ref{tab: rq4} presents the Mean Activation Value (MAV) of comment concepts across the layers of the LLMs for 10 SE tasks. Tasks are sorted by how often they yield the highest MAV for a concept in an LLM, followed by how often they yield the lowest. In 5 out of 12 experiments (3 LLMs $\mathnormal{\times}$ 4 concepts), code summarization results in the highest MAV, while in 6 out of 12 cases, code completion yields the lowest MAV. This demonstrates a clear difference in how comment concepts are activated across LLMs when the same code is used in different SE tasks.

Among the LLMs, both the Code LLM and the Reasoning LLM show their strongest activation for the Comment and Inline Comment concepts on the same task. In contrast, Javadoc is the only concept where LLMs diverge in the task that triggers their highest activation. Additionally, the Code LLM consistently shows the highest MAV across concepts, while the Reasoning LLM shows the lowest, except in the case of the general Comment concept. Overall, the weakest concept activation occurs in the Reasoning LLM during program repair, whereas the strongest occurs in the Code LLM during code documentation.

\begin{boxH}
\textbf{RQ4 Answer:} Activation of comment concepts depends on the SE task. Code summarization leads to the strongest activation, while code completion activates most weakly. 
\end{boxH}





\begin{table*}[ht!]
\caption{Mean Activation of Concepts ($P_c(e)$) Across Layers of Three LLMs in Each Task. \includegraphics[width=1em]{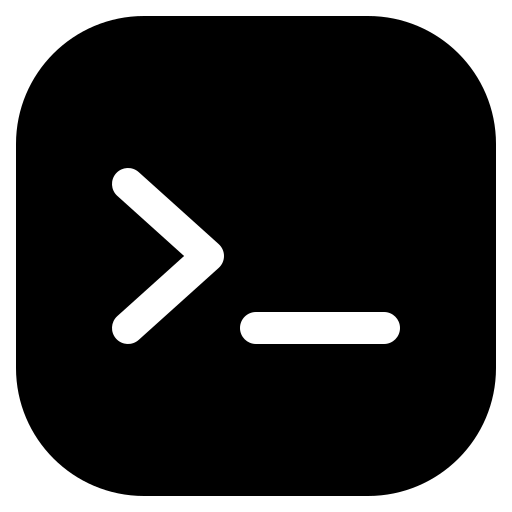} : Code LLM, \includegraphics[width=1em]{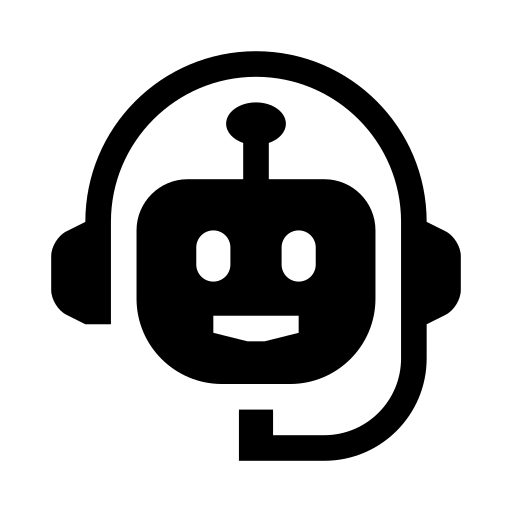} : Generic LLM, and \includegraphics[width=1em]{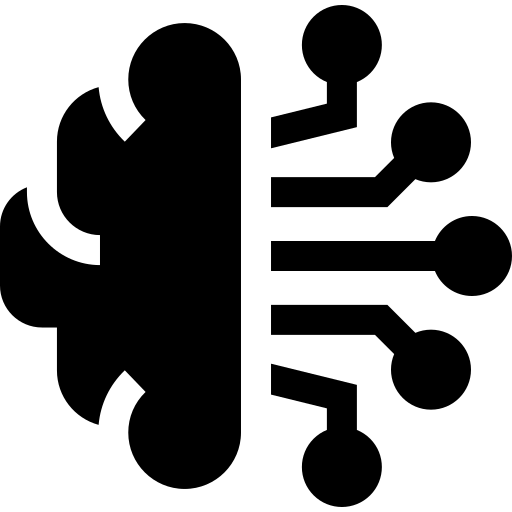} : Reasoning LLM. For each LLM, \highlighted{Value}: Largest mean activation of a concept. \underline{Value}: Smallest mean activation of a concept. $\uparrow$ is number of times a task gets the largest mean activation of concepts across LLMs $\downarrow$ is number of times it gains the lowest.}
\label{tab: rq4}
\centering
\resizebox{0.75\textwidth}{!}{
\begin{tabular}{l*{12}{|c}|cc}
\toprule
\multirow{2}{*}{\textbf{SE Task}} & \multicolumn{3}{c|}{\textbf{Comment}} & \multicolumn{3}{c|}{\textbf{Inline}} & \multicolumn{3}{c|}{\textbf{Javadoc}} & \multicolumn{3}{c|}{\textbf{Multiline}} & \multirow{2}{*}{$\uparrow$} & \multirow{2}{*}{$\downarrow$}\\
 & \includegraphics[width=1.35em]{terminal.png} & \includegraphics[width=1.35em]{chatbot.png} & \includegraphics[width=1.35em]{brain.png} & \includegraphics[width=1.35em]{terminal.png} & \includegraphics[width=1.35em]{chatbot.png} & \includegraphics[width=1.35em]{brain.png} & \includegraphics[width=1.35em]{terminal.png} & \includegraphics[width=1.35em]{chatbot.png} & \includegraphics[width=1.35em]{brain.png} & \includegraphics[width=1.35em]{terminal.png} & \includegraphics[width=1.35em]{chatbot.png} & \includegraphics[width=1.35em]{brain.png} \\
\midrule
Code Summarization &0.7 & \underline{0.25} & 0.63 &\cellcolor{rowgray}\textbf{0.79} & 0.58 & \cellcolor{rowgray}\textbf{0.64} &0.8 & 0.5 & \cellcolor{rowgray}\textbf{0.47} &0.56 & \cellcolor{rowgray}\textbf{0.55} & \cellcolor{rowgray}\textbf{0.51} & 5 & 1 \\
Code Review &\cellcolor{rowgray}\textbf{0.83} & 0.49 & \cellcolor{rowgray}\textbf{0.65} &0.78 & \cellcolor{rowgray}\textbf{0.72} & 0.38 &0.82 & 0.47 & 0.41 &0.5 & 0.47 & 0.23 & 3 & 0 \\
\midrule
Code Documentation &0.75 & 0.45 & 0.58 &0.54 & 0.51 & 0.38 &\cellcolor{rowgray}\textbf{0.86} & 0.45 & 0.35 &0.32 & 0.3 & 0.18 & 1 & 0 \\
Fault Localization &0.77 & 0.45 & 0.62 &\underline{0.36} & 0.59 & 0.51 &0.35 & \cellcolor{rowgray}\textbf{0.51} & 0.44 &0.27 & 0.37 & 0.21 & 1 & 1 \\
Test Generation &0.79 & 0.49 & \underline{0.55} &0.62 & 0.67 & 0.38 &0.6 & 0.46 & \underline{0.34} &\cellcolor{rowgray}\textbf{0.61} & 0.44 & 0.29 & 1 & 2 \\
Vulnerability Detection &0.79 & \cellcolor{rowgray}\textbf{0.53} & 0.64 &\underline{0.36} & 0.61 & 0.5 &\underline{0.33} & 0.48 & 0.43 &0.43 & 0.4 & 0.38 & 1 & 2 \\
\midrule
Code Refactoring &0.8 & 0.36 & 0.58 &0.57 & 0.62 & 0.37 &0.63 & 0.4 & 0.36 &0.38 & 0.29 & 0.21 & 0 & 0 \\
Program Repair &0.75 & 0.33 & 0.6 &0.49 & 0.59 & 0.39 &0.5 & 0.43 & 0.38 &0.29 & 0.31 & \underline{0.16} & 0 & 1 \\
Code Translation &0.75 & 0.42 & 0.58 &0.51 & 0.6 & 0.49 &0.56 & \underline{0.39} & 0.34 &0.51 & 0.46 & 0.22 & 0 & 1 \\
\midrule
Code Completion &\underline{0.67} & 0.43 & \underline{0.55} &0.44 & \underline{0.49} & \underline{0.36} &0.52 & 0.47 & 0.4 &\underline{0.26} & \underline{0.25} & 0.2 & 0 & 6 \\
\bottomrule
\end{tabular}
}
\end{table*}


\section{Implications} \label{implications}
\textbf{For Researchers.}
Our study represents a first step toward analyzing LLMs’ reliance on an internalized code concept, namely comments. 
By identifying that comments are internalized as distinct latent concepts, our study advances the broader understanding of how LLMs reason about code. It reveals that LLMs do not simply process source code as raw text, but instead decompose it into semantically meaningful components, such as code, comments, and structural cues, that interact to shape reasoning and decision-making. Understanding these internal mechanisms helps bridge the gap between surface-level input effects and the model’s internal cognitive processes, providing a foundation for more interpretable and controllable code reasoning models. Building on these insights, future research can identify which internalized code concepts most strongly affect model performance and develop methods to strengthen their representations. Conversely, recognizing code concepts that are poorly internalized or systematically detrimental could guide targeted interventions through unlearning \mbox{\cite{NEURIPS2024_171291d8}} or fine-tuning strategies.

Our findings from RQ3 indicate that, regardless of presence of comments in the input code, the internalization of comments as distinct concepts within the latent space of LLMs functions as a double-edged sword. Activating or suppressing this internal concept can substantially influence model performance across downstream software engineering tasks. While pinpointing the exact conditions under which this internalization helps or harms model behavior remains an open question, our results reveal a critical phenomenon that merits focused investigation into how such internalized comment concepts influence and shape LLMs’ performance in SE tasks.

Comment concepts are not uniformly triggered by the same code across tasks. Code summarization and code review lead to the strongest activations, while code completion consistently yields the weakest. This reinforces that task framing modulates the cognitive pathways LLMs use, even when the code input is identical. Understanding which tasks activate certain concepts can inform prompt design and dataset curation to elicit desired model behaviors.

\textbf{For Developers.} 
Our results show that comment concepts are most strongly activated in the code-specialized LLM, suggesting it is best suited for scenarios where developers aim to make full use of comments. However, activation patterns vary across LLMs and comment types. For example, the Inline Comment concept showed the highest activation in the generic LLM, indicating that such comments may be more effective for that LLM. Coupled with our RQ3 findings, developers should avoid over-relying on any single type of comment. For instance, while deactivating Javadocs improved the Generic LLM’s performance in code completion, it negatively affected its performance in code refinement. 
Overall, the effectiveness of comments in prompting desirable LLM behavior depends on the specific model, task, and comment type involved.
These findings suggest that developers should include a variety of comment types such as Javadocs, inline, and multiline comments, when documenting their code. Since each LLM may internalize and depend on these comment types differently, developers cannot reliably predict which type will most effectively support a model's understanding or reasoning without empirical evaluation. By maintaining comment diversity, developers can provide richer contextual cues that improve model comprehension across different scenarios. 
Furthermore, practitioners and researchers can assess which comment types a specific model relies on most using our proposed approach, paving the way for more intentional documentation practices tailored to model behavior.
Lastly, our findings reveal that code comments, long recognized as essential for code comprehension and software maintenance \mbox{\cite{xia2018pc, jabrayilzade2024taxonomy, gene2022forester}}, remain equally critical in the era of LLM-assisted software development. Consequently, maintaining both the quality and adequacy of comments continues to be an important concern for development teams.



\textbf{For Tool Builders.} 
Tool builders should design LLM-powered systems to be both task-aware and concept-sensitive. Since our results show that LLMs activate comment concepts differently depending on the task, tools should dynamically adapt how comments are used or generated based on the specific task at hand. Additionally, different LLMs respond more favorably to certain types of comments-for instance, inline comments are more effective in generic LLMs, while Javadocs are more robustly recognized by code-specialized models. This suggests that tailoring prompts or annotations to align with each model’s conceptual strengths-for example, injecting concise inline comments when using a generic LLM, or leveraging rich documentation-style comments when using a code-tuned model-can improve performance. Tools can incorporate this insight by automatically formatting or rewriting comments during pre-processing to match the preferred comment type for the underlying model. Finally, the ability to manipulate comment concepts through activation or deactivation suggests a promising avenue for tool builders: they can tune model behavior for a given task by selectively modifying conceptual inputs rather than retraining the model or altering the raw code, enabling lightweight debugging and targeted performance boosts.

\section{Threats to Validity} \label{threats}
\textbf{Construct Validity} concerns whether our methods accurately capture the internalization of comment concepts in LLMs. One threat is the potential misrepresentation of concepts in the training data for linear classifiers. To mitigate this, we curated a separate dataset from top-starred Apache Java repositories and applied strict filtering to ensure alignment with defined comment types (e.g., Javadocs, inline). Another concern is classifier overfitting; we addressed this by evaluating performance across varying training sizes and layers. Finally, while our use of CAV assumes linear separability, prior work supports this assumption \cite{kim2018cav, xu2024safetyrisks}, and our empirical results further validate its applicability in our setting.

\textbf{Internal Validity} relates to whether observed performance changes in LLMs are due to the perturbation of comment concept activations rather than external factors. One threat is confounding input variations, such as prompt phrasing or code properties. To mitigate this, although we adopted three different benchmarks that query LLMs with different pieces of code in RQ3, when answering RQ4 we used the same code base across tasks and adopted minimal zero-shot prompts to capture task semantics without introducing variations in form. To avoid unintended interference from multi-layer perturbations, we limited interventions to layers where classifier accuracy exceeded a threshold-ensuring perturbations target only layers where the concept is reliably encoded. We computed minimal $\epsilon$ values to apply the smallest necessary perturbation, as prescribed in prior work \cite{xu2024safetyrisks}, reducing the likelihood of side effects. All inference was performed with the temperature set to 0 to ensure determinism and eliminate randomness in outputs. Lastly, to mitigate the risk of data leakage, a common concern in LLM studies in SE \cite{zhou2025lessleak}, we operated at the representation level rather than the input level. By applying CAV-guided perturbations to internal activations instead of querying the model with specific examples, we isolate the influence of learned concept representations from potential memorization. This design reduces reliance on surface-level patterns, making our findings more robust and reflective of genuine conceptual internalization. 

\textbf{External Validity} concerns the generalizability of our findings. A key threat to generalizability is the limited architectural diversity, as all evaluated models are variants of Qwen2.5. Although these span code-focused, general-purpose, and reasoning-oriented variants, the results may not directly generalize to other LLM families, such as LLaMA \cite{dubey2024llama}. To mitigate this, we evaluated three distinct SE tasks using a diverse set of metrics. Moreover, using real-world code from Apache projects to construct classifier training data helps ensure that concept boundaries reflect practical coding practices rather than benchmark-specific artifacts. While our experiments focus on Java, the proposed methodology is language-agnostic and can be extended to other programming languages.

\section{Conclusion} \label{conclusion}
Our study presents the first concept-level interpretability analysis of LLMs in SE, revealing that models not only internalize comments as distinct latent concepts but also differentiate between subtypes such as Javadocs, inline, and multi-line comments. Using CAVs, we demonstrate that these internalized concepts are functionally significant. Manipulating these concepts in the embedding space produces substantial, task-dependent performance shifts across SE tasks. Notably, comment reliance varies widely-code summarization strongly activates comment representations, while code completion shows minimal sensitivity. These findings suggest that internal concepts are more than passive representations; they actively shape model behavior. This opens up new opportunities for building steerable SE tools that leverage concept-level manipulation. Future work should expand this investigation to higher-level software constructs (e.g., modularity, testability, API usage), enabling deeper insight into how LLMs reason about software and guiding the development of more interpretable and controllable models. All code and data are available in the supplementary material \cite{SupplementaryMaterials}.
\printbibliography










\end{document}